\documentclass[a4paper]{article}

\newcommand{\xmax}{\ensuremath{X_{\rm max}}}
\newcommand{\lsim}{\mathrel{\hbox{\rlap{\lower.55ex \hbox{$\sim$}} \kern-.3em \raise.4ex \hbox{$<$}}}}
\newcommand{\gsim}{\mathrel{\hbox{\rlap{\lower.55ex \hbox{$\sim$}} \kern-.3em \raise.4ex \hbox{$>$}}}}

\usepackage{icrc2013}

\title{Optimization of the Orbiting Wide-angle
Light Collectors (OWL) Mission for
Charged-Particle and Neutrino
Astronomy}

\shorttitle{Optimization of the OWL Mission}

\authors{
John F. Krizmanic$^{1,2}$,
John W. Mitchell$^{2}$,
\& Robert E. Streitmatter$^{2,3}$
for the OWL Collaboration.
}

\afiliations{
$^1$ CRESST/Universities Space Research Association \\
$^2$ NASA Goddard Space Flight Center, Greenbelt, Maryland 20771 USA \\
$^3$ retired \\
}

\email{john.f.krizmanic@nasa.gov}

\abstract{OWL \cite{OWL2004} uses the Earth's atmosphere as a vast calorimeter to fully enable the emerging field of charged-particle astronomy with high-statistics measurements of ultra-high-energy cosmic rays (UHECR) and a search for sources of UHE neutrinos and photons. Confirmation of the Greisen-Zatsepin-Kuzmin (GZK) suppression above $\sim 4 \times 10^{19}$ eV suggests that most UHECR originate in astrophysical objects. Higher energy particles must come from sources within about 100 Mpc and are deflected by $\sim 1$ degree by predicted intergalactic/galactic magnetic fields. The Pierre Auger Array, Telescope Array and the future JEM-EUSO ISS mission will open charged-particle astronomy, but much greater exposure will be required to fully identify and measure the spectra of individual sources. OWL uses two large
telescopes with 3 m optical apertures and 45 degree FOV in near-equatorial orbits. Simulations of a five-year OWL mission indicate $\sim 10^{6}$ km$^2$ $\cdot$ sr $\cdot$ yr of exposure with full aperture at $\sim 6 \times 10^{19}$ eV. Observations at different altitudes and spacecraft separations optimize sensitivity to UHECRs and neutrinos. OWL's stereo event reconstruction is nearly independent of track inclination and very tolerant of atmospheric conditions. An optional monocular mode gives increased reliability and can increase the instantaneous aperture. OWL can fully reconstruct horizontal and upward-moving showers and so has high sensitivity to UHE neutrinos. New capabilities in inflatable structures optics and silicon photomultipliers can greatly increase photon sensitivity, reducing the energy threshold for $\nu$ detection or increasing viewed area using a higher orbit. Design trades between the original and optimized OWL missions and the enhanced science capabilities are described.}

\keywords{Cosmic Rays, Extensive Air Showers, UHECRs, Space-based observation, Neutrino, OWL}

\begin{document}
\maketitle

\begin{figure}
\centering
  \includegraphics[width=.40\textwidth]{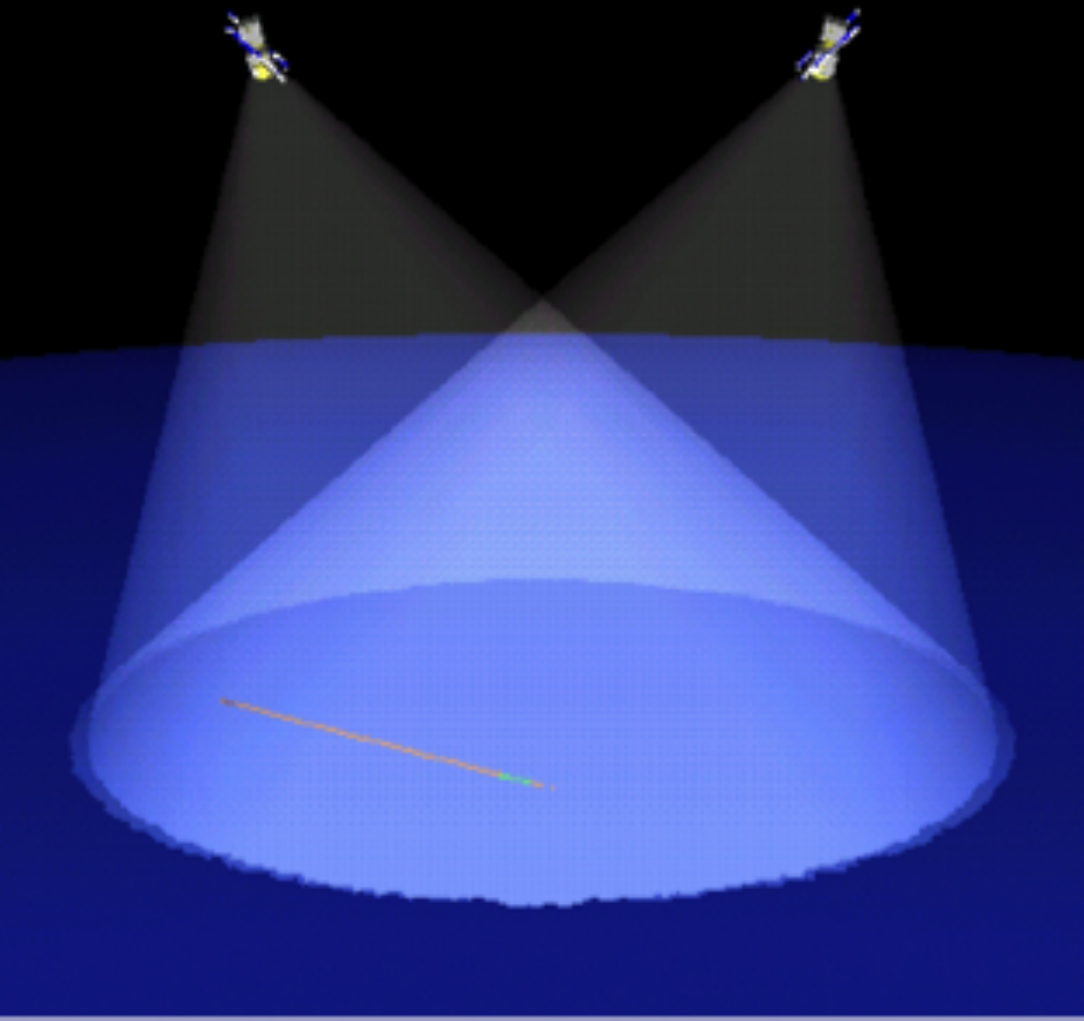}
  \caption{The OWL concept: two `eyes' stereoscopically viewing an extended air shower from low-Earth orbit. The FOV and  common atmospheric volume are highlighted.}
  \label{owlconcept_fig}
\end{figure}

\section{OWL Measurement Technique}
Excitation of atmospheric nitrogen molecules by particle showers produces a UV luminous disk, a few meters thick and $\sim$100 m wide moving close to c. OWL uses a fast, highly pixelized camera (eye) to resolve shower spatial and temporal development with monocular and stereo modes. In monocular, an individual camera images the projection of the shower onto a plane normal to the viewing direction. Distance from the camera must be resolved using other information. Differential arrival times of photons along the shower track give the angle of the shower relative to the viewing plane. The Cherenkov spot where tracks intersect the ground or clouds can establish distance, but this greatly reduces the aperture since tracks with shallow inclination may leave the viewing area. Stereo observation completely resolves all three spatial dimensions and and has the crucial advantage that atmospheric light absorption or scattering can be corrected. The fast timing for monocular operation provides supplementary track information in stereo.

\vspace{-3mm}

\section{OWL  Instrument and Mission}
The OWL instrument and mission were developed through extensive NASA studies. Challenges included the large optical collecting power needed to image particle showers from orbit and the difficulty in packaging the optical system in a standard launch vehicle shroud. The optical system itself has very modest requirements compared to astronomical telescopes and is a `ight bucket' with performance requirements closer to a microwave dish. Showers are at most several tens of kilometers in length, and the natural measurement scale is about 1 kilometer. The corresponding optical angular resolution at 1000 km is $\sim$1 milliradian, over 4 orders of magnitude larger than the diffraction limit.

OWL is an f/1 Schmidt camera with a 45$^\circ$ full FOV. The deployable primary mirror is 7.1 m in diameter, the corrector is 3.0 m, and the focal plane is 2.3 m. A bandpass filter on the corrector suppresses background. The effective aperture is 3.4 m$^2$. The primary is a lightweight composite with a central octagonal section and eight petals. The satellites launch with the petals folded upward, the corrector collapsed on the focal plane, some shield material in a storage volume, and the shutter closed. After deployment the petals are aligned using actuators to focus a point light source located at the center of curvature of the primary. The entire optical system is covered by an inflatable light and micrometeroid shield and closed out by a redundant shutter system. The shield with inflatable support ring and strengthening/shaping ribs can be rigidized. A UV laser for atmospheric characterization is located at the back of the focal plane and fires through the center of the corrector to a small steering mirror system. Laser light reflected by clouds is detected and measured using the OWL focal plane.

\begin{figure}
  \centering
  \includegraphics[width=3.0in]{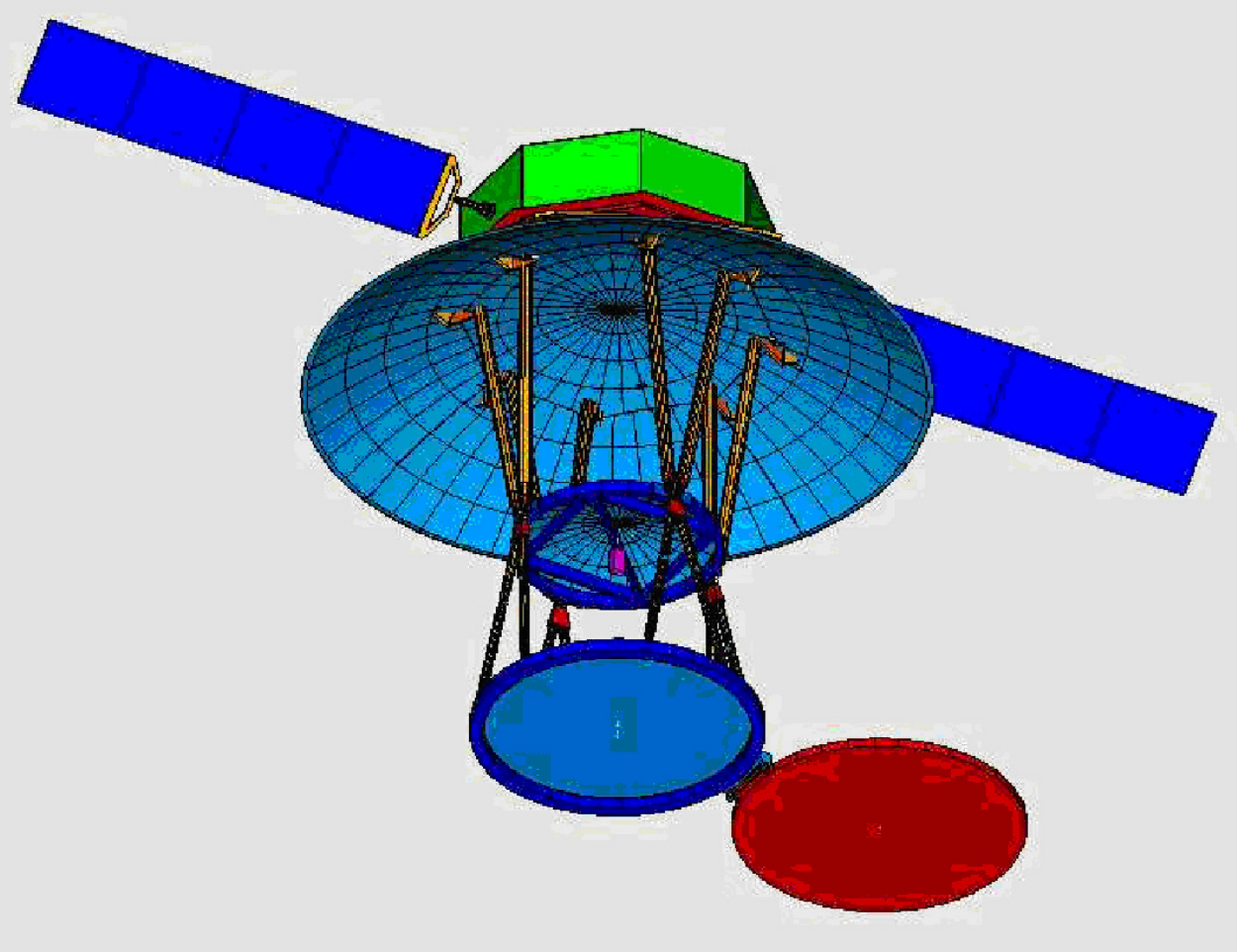}
\caption{Fully deployed OWL instrument with light shield removed for clarity.}
  \label{fig:figa}
\vspace{-3mm}
\end{figure}

The 4.9 m$^2$ focal plane is divided into $\sim$5 10$^5$ pixels that must measure single photoelectrons (pe), requiring high gain photodetectors. The baseline uses multianode vacuum PMTs but UV sensitive silicon PMTs (UVSiPM) currently in development would simplify the focal plane, reduce its mass, and double quantum efficiency. The natural time scale for readout is $\sim$1 $\mu$s since showers perpendicular to the viewing direction cross pixels in 3.3 $\mu$s. Sub-pixel readout of 0.1 $\mu$s is used to improve reconstruction systematics and angular resolution, and to support monocular operation. On each pixel, a switched capacitor array (SCA), with 3000 cells switching at 10 MHz, acts as an analog ring buffer to continuously store data for 300 $\mu$s, 90 pixel crossing times. Triggers use hierarchical hardware and software algorithms to suppress background and localize shower tracks on the focal plane. First level triggers are generated by local electronics based on spatial and temporal signal development around each pixel using automatic adaptive thresholds to compensate for background light. Look-up tables identify the region of pixels to read out. SCAs in that region are halted after 200 $\mu$s, recording 30 pixel crossing times before the trigger and 60 after, so the trigger can be generated at any point from the start of a shower to its peak. Events are selected within 2 sec of triggers, and fast track determination enables the LIDAR laser to slew and scan. A separate fast trigger detects upward-going neutrino events when the satellites are separated less than the width of the Cherenkov light pool. The trigger rate is $\sim$35 Hz and $<3\%$ of the focal plane is read in each event.

The OWL satellites are launched as a dual manifest into $\sim$10$^\circ$ circular orbits initially at 1000 km. The stereo viewed surface area is $\sim 10^{6}$ km$^2$, giving an instantaneous UHECR aperture of $\sim 2 \times 10^6$ km$^2$ sr. An initial separation of 10-20 km for $\sim$3 months is used to search for neutrinos passing through the Earth and initiating upward-going showers. The spacecraft then separate to 500 km for $\sim$2.5 years to measure the highest-energy UHECR. Then, altitude is decreased to 600 km  to reduce the detection threshold and measure lower energies. The lead instrument is pitched back (14 degrees from nadir at 1000 km), and the trailing instrument is pitched forward. Each instrument is completely shuttered whenever it might be exposed to direct or reflected sunlight or significant moonlight. The instruments are completely independent and do not require space-to-space communication. Data are combined on the ground using GPS time stamps.  All data are telemetered.  After event location and crude track determination, a series of laser shots are automatically taken along the track by each instrument to determine local atmospheric conditions. In addition, a random scan of the FOV is made at approximately one laser shot per second to characterize the average cloud obscuration. This is complemented by data obtained from IR-imaging satellites.

\begin{figure}[t]
  \centering
  \includegraphics[width=0.45\textwidth]{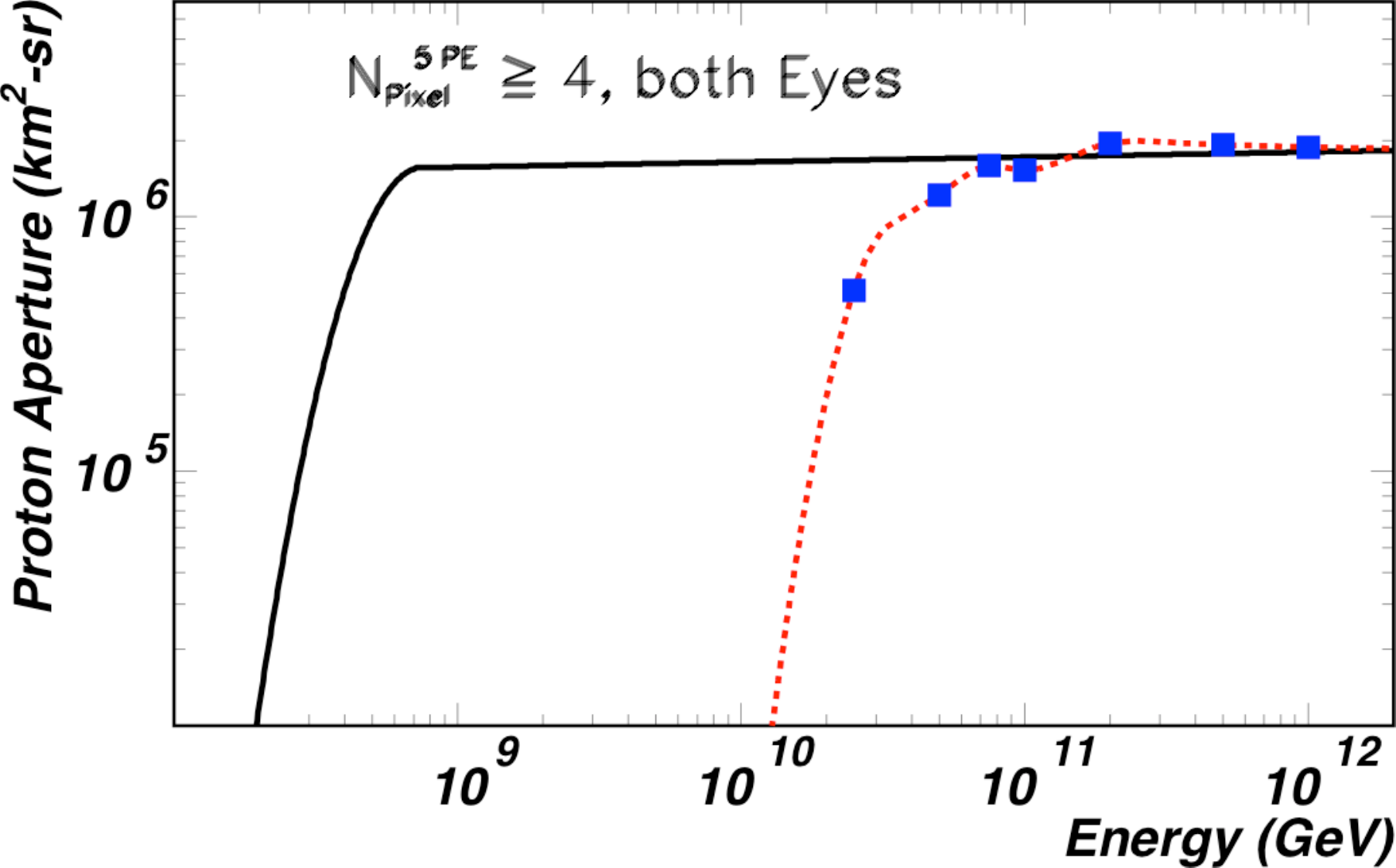}
\vspace{-1mm} 
  \caption{OWL and GreatOWL instantaneous proton aperture.}
  \label{CRaperture}
\vspace{-3mm} 
\end{figure}

\vspace{-3mm}

\section{GreatOWL Scientific Motivation}

OWL employs fully mature technologies and its performance has been extensively studied via simulations to show that the needed exposure for UHECR's above $10^{19}$ eV, was achieved. Figure \ref{CRaperture} shows the OWL UHECR aperture of nearly $10^{6}$ km$^2$ sr compared to an enhanced `GreatOWL'.  Studies indicate an \xmax\ resolution for stereo reconstructed events of $\sim 60$ g/cm$^2$ around $10^{20}$ eV improving to $\sim 35$ g/cm$^2$ at $10^{21}$ eV \cite{Tareq}. Thus while OWL has enormous exposure, its ability to distinguish composition would be limited.  Its potential to perform UHE neutrino measurements has also been simulated, partially driven by the search for exotic astrophysical phenomena that could be detected (since severely constrained by ANITA \cite{ANITA}.) With a threshold energy $\gsim 10^{19}$ eV, the OWL has minimal ability to measure GZK-cosmogenic $\nu$'s. Upward Cherenkov signals from GZK $\nu_\tau$'s interacting in the Earth has also been evaluated \cite{Krizmanic1}, but while the observation energy threshold is significantly reduced to below $10^{17}$ eV, the predicted event rate based on the Bartol GZK $\nu$ flux model \cite{Bartol} is only $\sim 1$/year.

If the energy threshold could be lowered to $\lsim 10^{18}$ eV, the exposure to the GZK $\nu$ flux would increase significantly.  Furthermore, UHECR composition could be measured above $10^{19}$ eV since the photostatistics and characterization of shower development both improve.  Below we discuss the gains enabled by increasing the optics size by a factor of $\sim 6$ and improving the focal plane quantum efficiency by a factor of two for a net factor of 70. We use this increased performance to define that of GreatOWL.

\vspace{-1mm}

\subsection{Improvement to GZK Neutrino Sensitivity}

\begin{figure}[t]
  \centering
  \includegraphics[width=0.22\textwidth]{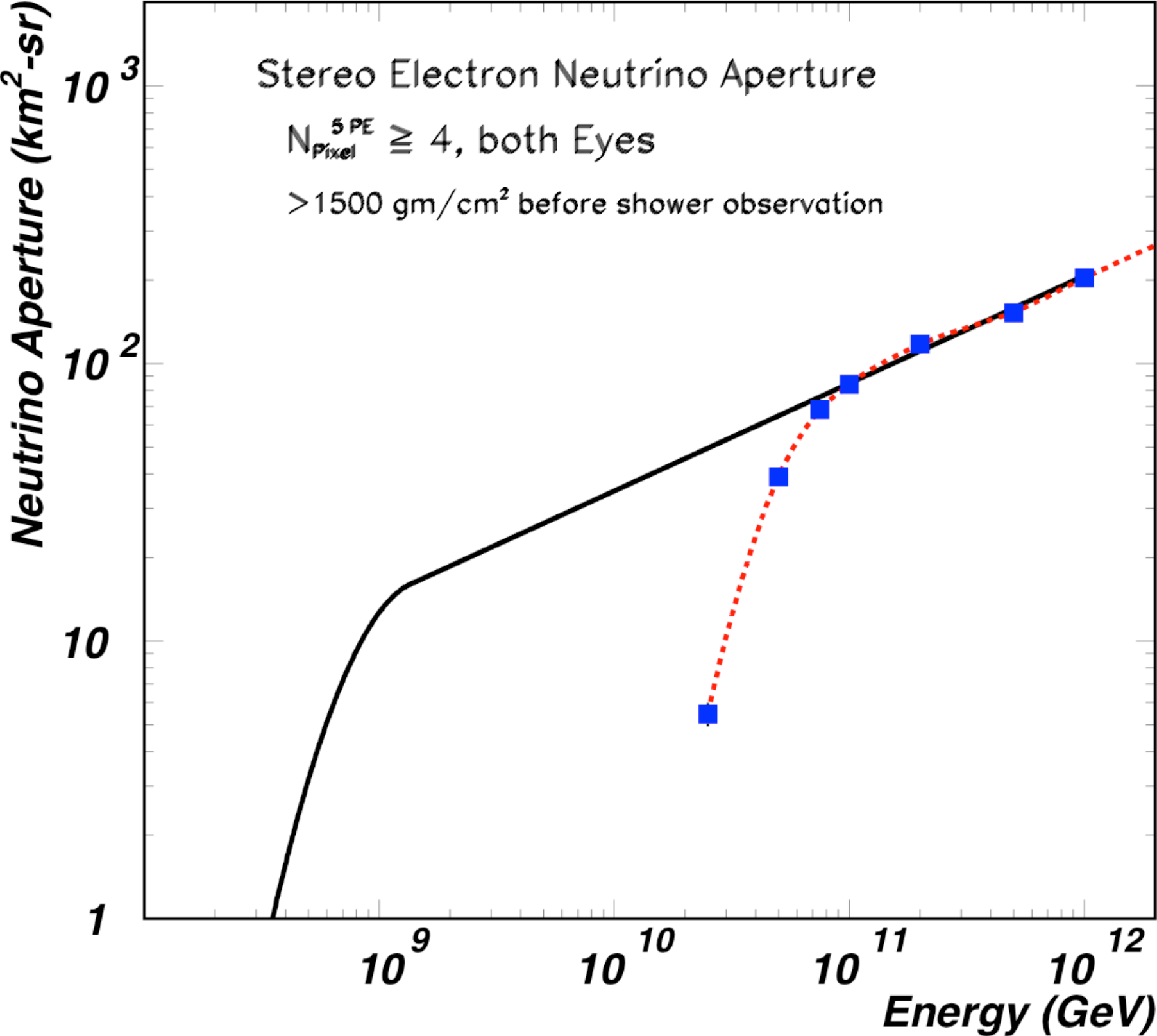}
 \hspace{1mm}
  \includegraphics[width=0.224\textwidth]{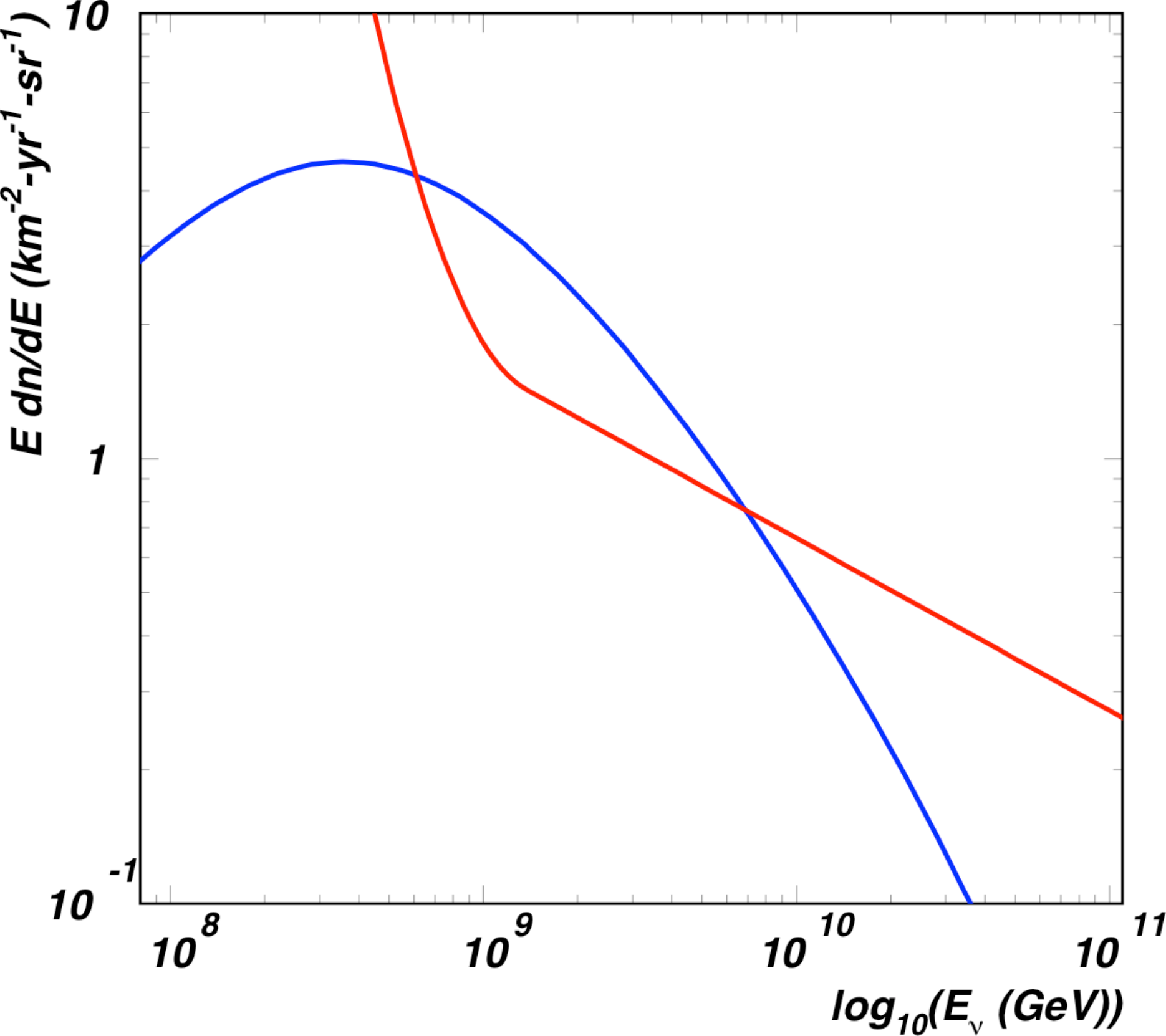}
\vspace{-1mm} 
  \caption{Left: OWL and GreatOWL instantaneous $\nu_e$ aperture. Right: OWL's extended 90\% confidence level $\nu_e$ limit compared to the $\nu_e$ Bartol oscillated GZK $\nu_e$ flux prediction.}
  \label{nuplots}
\vspace{-3mm} 
\end{figure}

The left panel in Figure \ref{nuplots} shows the instantaneous OWL and GreatOWL $\nu_e$ apertures based on modeling $\nu_e$ interactions in the Earth's atmosphere after requiring more than 5 photo-electrons in at least 4 temporal or spatial bins in each `eye' and the first portion of the observed $\nu$-induced air shower $> 1500$ g/cm$^2$ deep in the atmosphere.  Above the asymptotic aperture, the response increases as a power law with a spectral index very close to that of the $\nu$ cross section \cite{Quigg}, i.e. $\sim 0.363$. Since this holds for $\nu$ cross sections down to $\sim 10^{16}$ eV, extending the $\nu$ aperture to lower energies is reasonable. The 90\% confidence flux limit, based on the scaled aperture and 10\%  duty cycle is shown in the right panel of Figure \ref{nuplots} and is compared to the expected $\nu_e$ flux, which is determined by summing the muon and electron components of the Bartol GZK $\nu$ flux and dividing by 3 to account for $\nu$ oscillations. The GreatOWL flux limit is below the prediction between approximately $10^{17.8}$ eV and $10^{18.9}$ eV. 

Above $\sim 10^{16}$ eV, the lepton produced in both a charged- and neutral-current neutrino interaction carries the majority of the incident $\nu$ energy.  Assuming that 80\% is carried in the leptonic channel and 20\% in the hadronic channel, the $\nu$ aperture can be convolved with the Bartol-GZK flux prediction (assuming equal flavor oscillations) to yield an event rate prediction assuming a 10\% duty cycle.  One-year neutrino event rates are presented in Table \ref{nuTable} for the various $\nu$-interaction channels: $\nu_e$ CC interactions put 100\% of the incident $\nu_e$ energy into a shower at the interaction point, $\nu_\mu$ CC interactions put 20\% of the $\nu_\mu$ interaction into a shower at the interaction point (the muon is assumed not to be observed) but with a threshold energy $\times 5$ higher than that for a $\nu_e$ CC event, a $\nu_tau$ CC interaction puts 80\% into a shower (it is assumed that the `double bang' is not observed) but with a slightly higher energy threshold than $\nu_e$ CC events, and neutral current (NC) $\nu$ interactions mimic that for the $\nu_\mu$ interaction model, e.g. 20\% of the $\nu$ interaction into an air shower at the interaction point.  Simulations indicate that only $\sim 20\%$ of 'double bang' $\nu_\tau$ CC events will have the two spatially separate air showers in the OWL FOV at $5 \times 10^{18}$ eV.  For the $\nu_\tau$ CC events, a 15\% reduction in the predicted rate is applied to account for the $\tau \rightarrow \mu$ decays that are not observable.

\begin{table}[h]
\begin{center}
\begin{tabular}{|l|c|c|}
\hline Interaction Channel & Energy Threshold & Events/Year \\ \hline
$\nu_e$ CC  & $\sim 10^{18}$ eV  & 15 \\ \hline
$\nu_\mu$ CC  & $\sim 10^{18.7}$ eV  & 8 \\ \hline
$\nu_\tau$ CC  & $\sim 10^{18.4}$ eV  & 9 \\ \hline
$\nu_\lambda$ NC  & $\sim 10^{18.7}$ eV  & 10 \\ \hline
{\bf TOTAL} & & {\bf 42}  \\ \hline
\end{tabular}
\caption{The predicted channel-dependent neutrino event rate per year based on the GreatOWL $\nu$ aperture and the oscillated prediction of the Bartol GZK neutrino model and a 10\% duty cycle. The combined results yield 42 neutrino event per year or 210 neutrino events for a 5-year mission.}
\label{nuTable}
\vspace{-3mm}
\end{center}
\end{table}

As detailed in the table, over 200 GZK $\nu$ events could be observed by lowering the energy threshold to $\lsim 10^{18}$ eV, based on the (proton-dominated) Bartol model. Recent calculations \cite{Olinto} have evaluated the impact of a heavy UHECR nuclear composition on the GZK-cosmogenic neutrino flux.  If the $\nu$ flux reduction is only a factor of $\sim 10$ (as suggested by some of the models), then GreatOWL would still expect to observe $\sim 20$ GZK $\nu$ interactions in 5 years.  It is also important to note that if the duty cycle can be increased as suggested in \cite{Adams}, then the expected $\nu$ event rate would also increase.


\subsection{Improvement to \xmax\ Resolution}

A key component of \xmax\ resolution is pe statistics. To quantify the gain from increasing sensitivity by a factor of 70, we compare the effects of the pe statistics at $10^{20}$ eV from the OWL Monte Carlo simulation and those anticipated in GreatOWL at the more physically interesting energy of $10^{19.5}$ eV. For OWL at $10^{20}$ eV, the pe distribution peaks at around 150.  The altitude of shower maximum for these events is $\sim 8$ km, leading to an approximate sampling of 20 g/cm$^2$ based on a 1 $\mu$s signal integration time. 

\begin{figure}[t]
  \centering
  \includegraphics[width=0.235\textwidth]{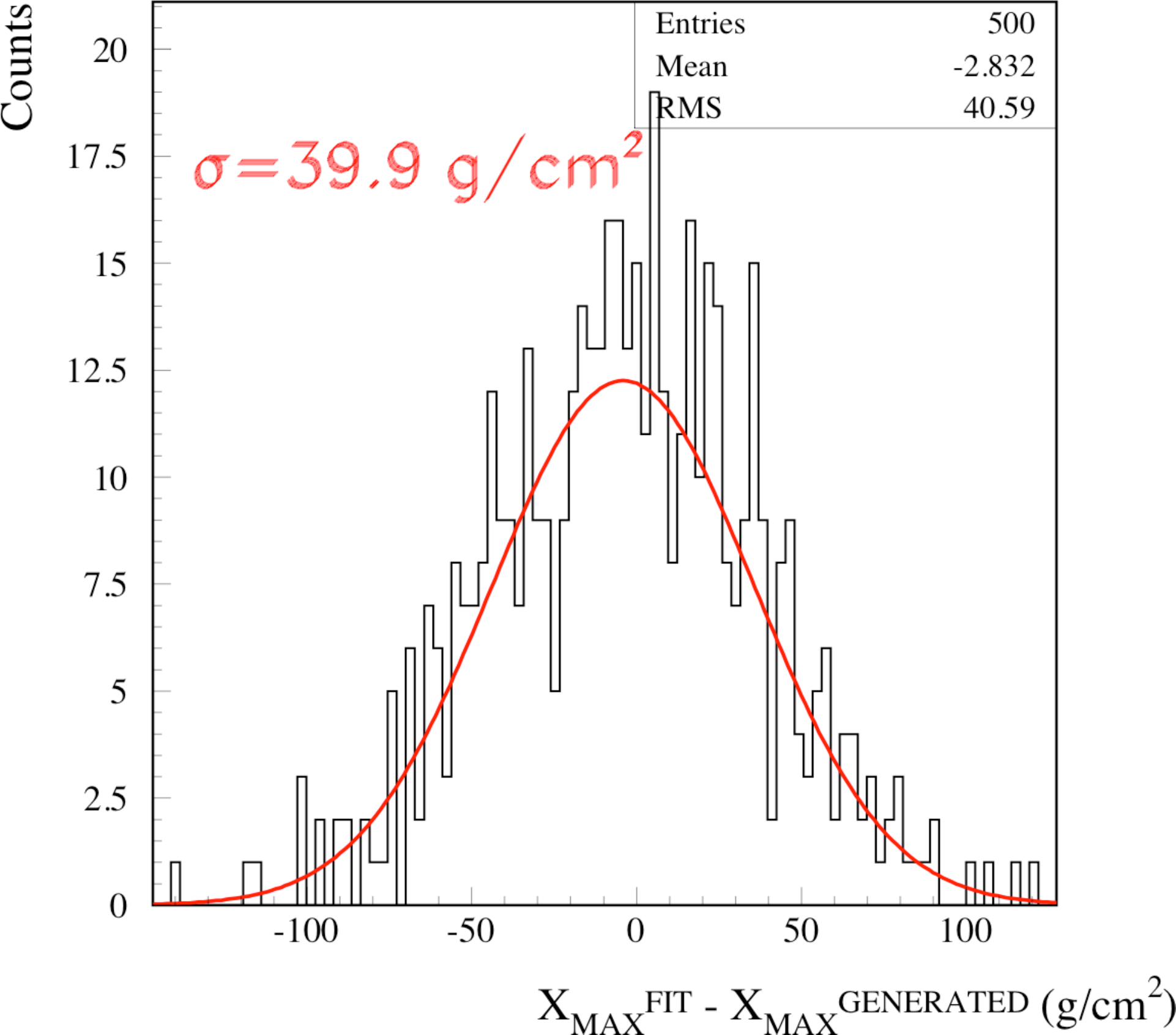}
  \includegraphics[width=0.235\textwidth]{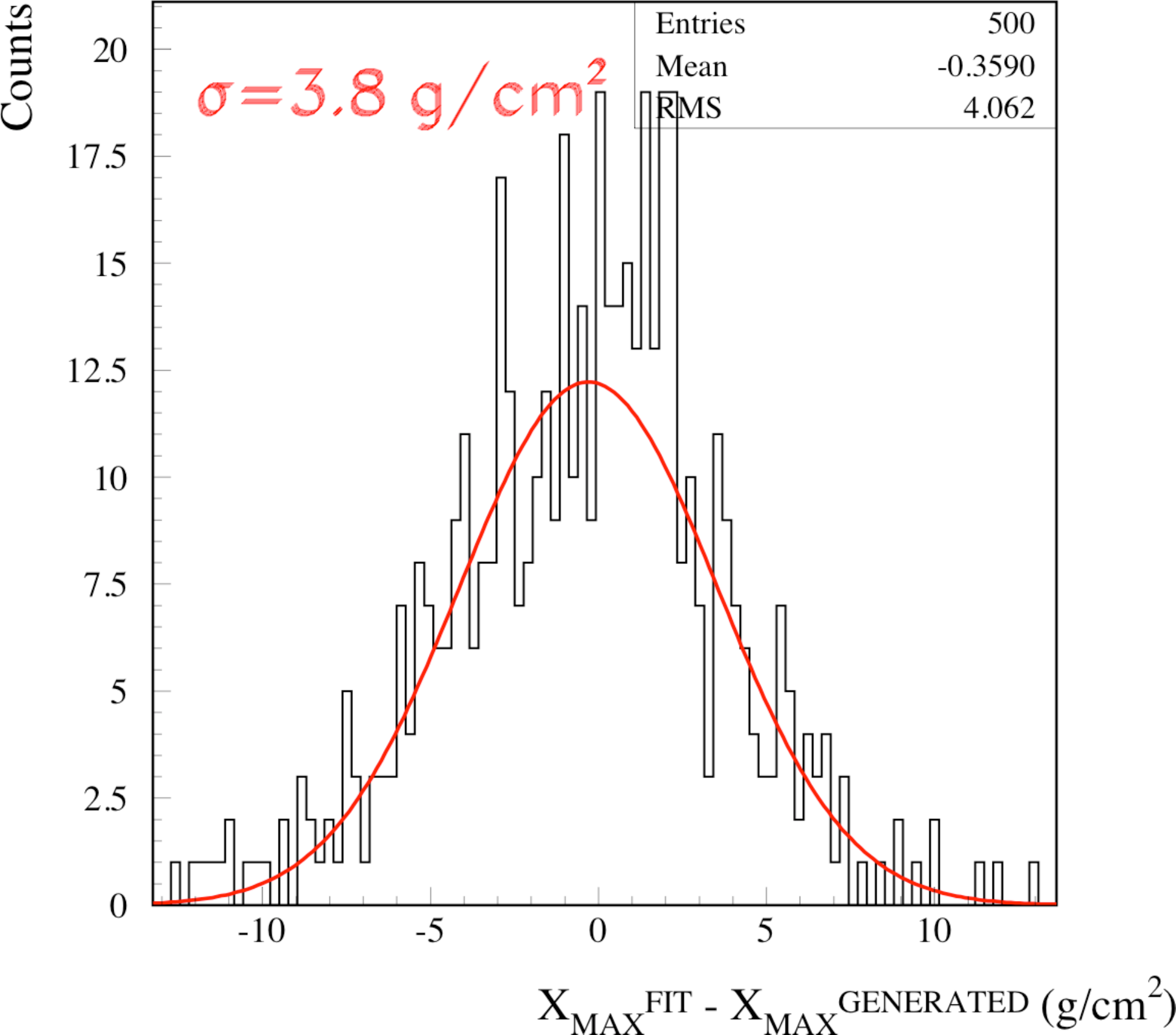}
\vspace{-1mm} 
  \caption{Left: The \xmax\ resolution at $10^{20}$ eV based on only pe-statistics for OWL. Right:The \xmax\ resolution at $10^{19.5}$ eV based on only pe-statistics for GreatOWL.}
  \label{OWLxmax}
\vspace{-3mm} 
\end{figure}

Gaisser-Hillas (GH) functions are used to generate air shower profiles with 20 g/cm$^2$ bins and scaled to have bin-summed contents corresponding to the expected signal: 150 pe for OWL at $10^{20}$ and $\sim 3200$ pe for GreatOWL at $10^{19.5}$ eV.  The bin-by-bin expectation is Poisson fluctuated, and the distributions are fit using a GH parameterization.  The fit \xmax\ is subtracted from the generated \xmax\ and the distribution is fit to a Gaussian to obtain the standard deviation.  The procedure uses independent air shower GH parametrized events at the appropriate energy, and the procedure is repeated 500 times for each distribution.

The left plot in Figure \ref{OWLxmax} shows the resulting distributions for OWL and exhibits a \xmax\ resolution of 40 g/cm$^2$ {\it based on pe statistics alone}. Thus even in the case of perfect event reconstruction with minimal systematic errors, the \xmax\ resolution cannot be better than 40 g/cm$^2$ at $10^{20}$ eV. The right plot in Figure \ref{OWLxmax} shows the resulting distribution for GreatOWL with an \xmax\ resolution of 4 g/cm$^2$. Thus if event reconstruction can be accomplished with minimal systematic errors, the \xmax\ resolution {\it could} be sufficient, e.g. $\sim 20$ g/cm$^2$ at $10^{19.5}$ eV and thus allow accurate composition measurements in the $10^{19}$ eV decade.  The fact that more of the air shower profile will be sampled by GreatOWL and the small angular pixel size (0.06$^\circ$) along with the 100 ns sampling time of the OWL design should aid the reduction of the effect of reconstruction systematics on the \xmax\ resolution, but these need to be quantified in a full Monte Carlo simulation coupled to robust reconstruction algorithms.  The point in this pe-focused study is that GreatOWL can {\it potentially} achieve sufficient \xmax\ resolution to enable UHECR nuclear composition studies.

\vspace{-3mm}

\section{The GreatOWL Instrument and Mission}
GreatOWL would have a 42 m diameter primary mirror, 18 m corrector and 13.8 m focal plane. This would be prohibitively massive and complex using conventional methods so we have baselined inflatable structures rigidized using some of the several techniques developed for large-area antenna and solar power systems. For OWL, we investigated rigidizable inflatable structures and adopted them for the light shield but not for the optical system. For GreatOWL, this is the only practical approach, and we outline a potential design. The primary mirror, the corrector, the struts forming the optical component supports, and the light shield will all be inflatables. The focal plane will be made partly using flexible circuit boards that can be folded on the sides of the spacecraft and deployed by inflatable gussets and ribs.

There are several other differences between OWL and GreatOWL. Because the UVSiPM can be exposed to significant light, especially with the bias off, a multi-layer light baffle/micrometeoroid shield will be used (similar to the approach used by JWST). There will be no shutter and when in sunlight, the instrument will be rolled to point toward deep space and the UVSiPM bias removed. As each spacecraft enters twilight, will be rolled to view the Earth and when near full dark (monitored by light sensors) the UVSiPM bias will be restored. The spacecraft will be in the optically dead area between the focal plane and the corrector to reduce loads on the optical struts during maneuvers. Heat will be piped to radiators on the deployment ring of the primary mirror, and heat pipes and wiring will run along one of the struts. Piecewise flexible photovoltaic (PV) panels will be located on the sides of the baffle. The octagonal spacecraft fits a 4.6 m shroud and contains all needed electrical and attitude control systems. With the UVSiPM there are no issues with interacting with magnetic torque bars as there would be with vacuum photomultipliers. Telemetry antennas on the spacecraft and will communicate through the dielectric material of the light baffles.

The GreatOWL reflector will be formed by an initially convex-convex inflatable with the reflecting surface on the inside. An inflatable ring at the edge of the mirror will be employed in the deployment sequence and will form the base for the light baffles. The spherical figure will be formed during inflation. After rigidization, the surface toward the spacecraft will be torn away by tethers on the ring and stowed permanently at the edge of the mirror outside the optical region.

The corrector is a refracting element and cannot easily be formed using only inflatable surfaces. Instead, the inflation will simply form a volume with the needed figure and transparent surfaces. This can then be filled with a polymer base and then catalyzed to form a solid lens. Alternatively, the volume can be filled with a transparent, low vapor pressure liquid. Small flexible membranes with surfaces normal to the optical axis or inflatable ribs can be added to establish the figure without significant optical impact. At launch, the corrector will be rolled and stowed on the side of the spacecraft outside the focal plane petals so the surface of the spacecraft opposite the focal plane is free for attachment to the launch vehicle.

The focal plane will be made of an octagonal center section of the same dimension as the spacecraft and and eight flexible `petals' folded onto its sides. The 1.8 cm pixels will be made of $6 \times 6$ UVSiPM arrays cooled to about 0$^\circ$ C to reduce dark noise. The readout electronics will follow original OWL design. The focal plane petals will be deployed using inflatable gussets and ribs attached to the sides of the spacecraft.

The instruments will be launched as a dual manifest, possibly on a relatively small vehicle such as a Falcon 9. For deployment, the primary mirror will first be moved away from the spacecraft by partly inflating its support struts. The support ring for the primary is then inflated to open the mirror to full diameter and the support struts are fully inflated. Next, the light baffles with the PV arrays and radiators. are deployed by inflating their extension tubes. The corrector is then moved laterally away from the spacecraft by inflating the support strut at the center of the packed material. When fully clear, the other two support struts are inflated to move the corrector into place. The primary mirror is then inflated and rigidized and covering surface is removed. Finally the corrector is inflated, rigidized and filled.

\vspace{-3mm}
\section{Conclusion}

We have detailed the significant improvement both GZK-cosmogenic neutrino measurement and potential for performing accurate UHECR composition measurements, if an enhanced GreatOWL can be built with the same FOV as OWL and an energy threshold $\lsim 10^{18}$ eV.  This requires OWL optics to be enlarged by a factor of $\sim 6$, in linear dimension, coupled with a focal plane with double the current quantum efficiency. Technologies such as inflatable optics and structures and UVSiPMs offer feasible solutions to these demanding requirements.  The fact that either version of OWL images the UV light from air showers $\sim 10^4$ away from the diffraction limit, and the tolerance to alignment errors is $\sim 1$ mm in the OWL designs supports the concept that inflatable structures can provide a cost-effective optical and mechanical solution.

\end{document}